\documentclass[pra,twocolumn,showpacs]{revtex4}
\usepackage{graphicx}
\usepackage{amsmath}
\def\Journal#1#2#3#4{{#1} {\bf #2}, #3 (#4)}

\newcommand{\n}{\nonumber}
\newcommand{\bn}{\begin{eqnarray}}
\newcommand{\en}{\end{eqnarray}}
\newcommand{\eml}{\end{multline}}
\newcommand{\bml}{\begin{multline}}
\newcommand{\h}{\hspace}

\begin{document}


\title{Thermodynamic and noise considerations
for the detection of microscopic particles in a gas by
Photoacoustic Raman spectroscopy }

\author{Kunal K. Das$^1$\footnote{Present Address:  Department of Physics, The
Pennsylvania State University, University Park, Pennsylvania
16802.\\
e-mail: kdas@phys.psu.edu}, Yuri V. Rostovtsev$^1$, Kevin
Lehmann$^2$, and Marlan O. Scully$^{1,3}$}
\affiliation{$^1$Department of Physics, Texas A$\&$M University,
College Station, Texas 77843}
\affiliation{$^2$Department of Chemistry, Princeton University,
Princeton, NJ 08544}
\affiliation{$^3$Departments of Chemistry and Aerospace and
Mechanical Engineering, Princeton University, Princeton, NJ 08544}

\date{\today}
\begin{abstract}
We develop a simple thermodynamic model to describe the heat
transfer mechanisms and generation of acoustic waves in
photoacoustic Raman spectroscopy by small particulate suspensions
in a gas. Using Langevin methods to describe the thermal noise we
study the signal and noise properties, and from the noise
equivalent power we determine the minimum number density of the
suspended particles that can be detected. We find that for some
relevant cases, as few as 100 particles per cubic meter can be
detected.
\end{abstract}


\maketitle

\section{Introduction}

Optoacoustics as a field has venerable roots dating back to 1880
when Alexander Graham Bell\cite{Bell} first found that periodic
illumination of colored substances generates sound.  The process
involves absorption of energy from a modulated light source by the
atoms or molecules in the medium, producing a variable heat source
which serves as a source for acoustic waves; the sound can be
detected by microphones or piezoelectric transducers.

Photoacoustic spectroscopy, as it is also known, has become a
valuable spectroscopic tool with extensive use in the study of
absorption spectra and electronic spectra of substances in both
condensed and gaseous phases \cite{Pao,cknrmp,tam,Rosencwaig}. The
type of atomic or molecular modes being studied determines the
frequency of the incident light used, which can be in the visible
range for electronic excitations or in the infra-red for
vibrational modes.

Photoacoustic techniques were first applied to Raman spectroscopy
in 1979 by Barrett and Berry \cite{Barrett}, to create the
nonlinear spectroscopic technique they called photoacoustic Raman
spectroscopy (PARS). It was subsequently applied to liquids
\cite{ckn} and to gaseous trace analysis\cite{Seibert}. Resolution
much higher than the natural linewidth of Raman transitions has
been achieved using narrow linewidth pulsed lasers \cite{Rotger}.

Numerous researchers in the last couple of decades have routinely
applied photoacoustic methods to determine structural and thermal
properties in solid state physics \cite{Mandelis,solidstate}, to
conduct non-intrusive studies of nanostructures \cite{nano}, to
analyze internal structure and dynamics of molecules in chemistry
\cite{chemistry}, and as a non-invasive way of characterizing
wavelength dependent optical properties of biological samples
\cite{Krebs}. But almost all such applications have involved
linear infrared excitations, and very often the photoacoustic
measurements were a supplement to separate Raman spectroscopic
analysis. In contrast there has been limited literature on the
usage of PARS as a tool by itself for spectral studies.

However as we will show here, PARS can become a very effective and
sensitive means of spectral detection of suspended solid
contaminants in air. The immediate motivation for our study arises
from the urgent necessity for reliable methods to detect trace
amounts of microbial spores like anthrax in the current context of
defense against bio-terrorism. Methods based on coherent Raman
spectroscopy are already being actively pursued \cite{fastcars}
where the spectroscopic signature of a characteristic molecular
constituent is sought. An optoacoustic method based on Raman
spectroscopy can take advantage of the same signature but offer an
alternative and experimentally simpler detection scheme. In fact
the first application of infra-red photoacoustic spectroscopy for
identifying bacterial spores was reported just a few months ago
\cite{Thompson} where the constituents in a mixture of diverse
specimen were identified correctly with $100\%$ accuracy.
Applications are viable also in a broader context since acoustic
detection of stimulated Raman emissions using the powerful lasers
available these days can lead to more sensitive trace analysis.

It is worth noting that application of photoacoustics to highly
sensitive detection of trace amounts of pollutants in gaseous
medium was originally responsible for its tremendous growth and
revival in the 1970's. Traditionally such applications have almost
always relied on infrared single photon absorption by the
pollutants. But using Raman scattering provides several
advantages: (i) the incident light can be in the visible range in
which both air and water are essentially transparent while they
absorb in the infrared; this is particularly relevant for
bacterial spores because most of the non-signature molecules are
water, (ii) adjusting the difference of the two input fields
allows tuning to different vibrational modes not available for
direct infrared excitation, and (iii) in stimulated Raman
scattering the signal depends on the product of two laser
intensities, and so it can be maximized using strong lasers.

Moreover new innovations in Raman spectroscopy, such as STImulated
Raman Adiabatic Passage (STIRAP) \cite{Bergmann} or using chirped
pulses allows complete population transfer from the ground state
to the first molecular vibrational state. Subsequent relaxation of
the excited state can generate sound waves. In fact experimental
evidence already exists \cite{harris}, that a coherently excited
molecular gas generates intense sound waves when the incident
electromagnetic fields are close to Raman two-photon resonance.

In this paper we will consider a model for a PARS experiment to
detect small ($\sim 1{\rm \mu m}$) suspended solid particle in
air.  We determine the signal strength by considering the basic
thermodynamic processes involved and describe the acoustic
disturbance by a wave equation. To describe the noise in the
system we use a Langevin method which has the advantage of being
able to include possible additional sources of noise with little
extra effort.  From the expressions for the signal and the noise
we determine the noise equivalent power (NEP) and from that the
minimum detectable particle density. We describe our model in Sec.
II, and then in Secs. III and IV we consider the thermodynamics of
the energy transfer from the fields to the photoacoustic cell. In
the following two sections V and VI we describe the acoustic waves
and evaluate the thermal noise and the NEP using second order
Langevin equations, and then in Sec. VII we bring all the elements
together to determine the minimum detectable number density for
the specific example of an anthrax spore.

\section{The model}

\begin{figure}\vspace{-2cm}
\includegraphics*[width=\columnwidth,angle=0]{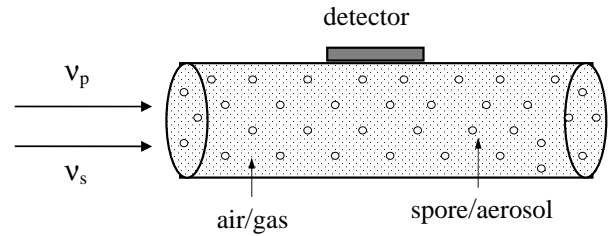}
\vspace{-7cm} \caption{Schematic configuration of our model for
optoacoustic detection of suspended particles (indicated by small
circles) in air contained inside a photoacoustic cell (shown as a
cylinder). The drive and the Stokes photons carry energies
$\hbar\nu_p$ and $\hbar\nu_s$.} \label{Fig1}
\end{figure}

The configuration we consider, shown in Fig.~\ref{Fig1}, is
typical of most photacoustic experiments where the driving lasers
pass through a gas/air-filled cylinder which may contain the Raman
active medium of interest in the form of suspended particles. The
detector is located at a position orthogonal to the direction of
propagation of the laser beams. Mirrors may be placed at the ends
of the gas cell for multiple passes of the lasers.

\begin{figure}
\includegraphics*[width=\columnwidth,angle=0]{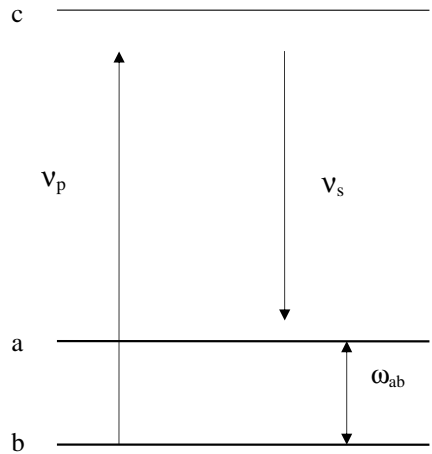}\vspace{-4cm}
\caption{Stokes-Raman scattering.  The driving field has frequency
$\nu_p$ and the Stokes field $\nu_s$, and they induce transitions
between levels $a$ and $b$ via an intermediate state $c$ which
could be virtual but may coincide with a real energy level.}
\label{Fig2}
\end{figure}

The Raman scattering needs to be Stokes-type because energy needs
to be deposited in the medium, and necessarily be stimulated in
nature because of the stronger signals produced. Thus there are
two input beams: the pump beam with energy $\hbar\nu_p$ and a
Stokes beam with energy $\hbar\nu_s<\hbar\nu_p$. The Raman
scattering scheme is shown in Fig.~\ref{Fig2} with $a$ and $b$
being the upper and the lower levels and energy difference is
$\hbar\omega_a-\hbar\omega_b=\hbar\omega_{ab}$.  Raman resonance
is achieved when $\nu_p-\nu_s=\omega_{ab}$.

Each of the spores we wish to detect has volume $V_S$ and contains
$N_S$ molecules; out of these a fraction $f_R$ are the Raman
active molecules we call $R$, and the remaining $1-f_R$ we take to
be predominantly composed of another species of molecule $W$,
usually water for a bacterial spore, that is effectively
non-absorbing in either of the input fields. The Raman scattering
deposits energy in the gas cell whereby heat is generated that
serves as the driving force for acoustic waves. The heat-energy
transfer happens in two steps:

\emph{Step 1}. The excitation energy $\hbar\omega_{ab}$ of the
molecules $R$ will be converted to heat energy via collisions
$R\leftrightarrow W$, $R\leftrightarrow R$ and $W\leftrightarrow
W$ among the molecules in a particle. This will depend on the
Raman scattering cross-section and the radiative and collisional
damping that happens \emph{within} a spore.

\emph{Step 2}.  The energy gained by molecules in the spore will
be transferred to the gas molecules in the tube.  Since the spore
is much larger than the gas molecules, this can be modelled by
kinetic theory.

If we define $H_R$ to be the energy absorbed \emph{due to Raman
scattering} per unit volume per unit time, then the energy
absorbed by a spore per unit time is $H_RV_S$. A fraction $\eta$
of this energy is converted to translational energy or heat of the
surrounding gas molecules within the time scales of acoustic
modes, and therefore that is the fraction available for the
acoustic signal. The spores will be randomly distributed in a gas
cell, and we define $\rho_S$ to be their average \emph{number
density}. The amount of energy available for acoustic signals per
unit volume of air per unit time is then given by:
\bn\label{power} H=\rho_S\times \eta\times H_R\times V_S\en
In order to determine the minimum detectable $\rho_S$, we will
have to find the minimum energy density $H$ that can produce a
detectable signal, and for this we first need to estimate all the
other quantities in the above expression as well as the noise
level.

\section{Raman induced heat energy}
In the Raman scattering process, the pump beam loses photons by
exciting the molecules and their subsequent de-excitation produces
Stokes photons, and the energy difference is deposited in the
medium. For each photon converted the deposited energy is
$\hbar\nu_p-\hbar\nu_s$. The total energy deposited is determined
by the number of photons \emph{gained} by the Stokes beam, which
has an exponential dependence on the interaction length $z$,
\bn n_s(z)=n_s(0)e^{g z}. \en
The gain coefficient $g$ is given by \cite{Yariv}
\bn
g=\frac{8\pi^2Nv_s^2}{\hbar\nu_s^3\delta\nu}\times\frac{d\sigma}{d\Omega}
\times I_p \times[1-e^{-\hbar(\nu_p-\nu_s)/(kT)}]\en
where $v_s=3\times 10^8{\rm ms^{-1}}/n(\nu_s)$ is the
\emph{velocity} of the Stokes photons with $n(\nu_s)$ being the
refractive index in the medium; $\delta\nu$ is the linewidth for
the Raman transition; $N$ is the \emph{density} of the active
medium. The last factor is the fractional population difference of
the upper and lower atomic levels for thermal occupation
probabilities. The gain is relatively small over the interaction
lengths $L$ which is on the scale of a spore-size for trace
occurrences, so we may write $e^{g_s L}\simeq 1+g L$. This also
means that the fractional change in the input beam intensities is
small so that we can treat the average values of $I_{p}$ and
$I_{s}$ as constants during the time of interaction. Thus the
change in the intensity of the Stokes signal is
\bn \Delta I_s=G I_pI_sL\en
where we defined the intensity-independent gain factor $G=g/I_p$.
By definition, intensity $\equiv$ power/area, therefore dividing
through by $L$ we have a relation for the rate of change of energy
of the Stokes beam per unit volume. Each photon converted deposits
energy $\hbar\nu_p-\hbar\nu_s$ in the medium, so the rate of
energy absorbtion per unit volume of the medium is given by
\bn \frac{\Delta I_s}{L}=\frac{(\nu_{p}-\nu_s)}{\nu_{s}}G
I_pI_s\en
There are three possible fates for this energy:  (i) it can be
lost by radiation, (ii) energy from one molecule can be
transferred to excite another molecule via collisions or (iii) the
energy will be converted to translational energy via collisions.
The last process is the one which will directly lead to heating of
the system. Since the excitations are vibrational in nature the
second process will also cause heating as well after a few
collisions. We denote the radiative decay rate by $\Gamma_r$ which
corresponds to the first process of loss, and we denote the
collisional loss by $\Gamma_c$ which corresponds to the last two
processes. With these considerations we can now relate the heat
absorbed per unit volume of the medium to the Raman gain factor
and the input laser intensities
\bn\label{source}
H_R=\frac{\Gamma_c}{\Gamma_c+\Gamma_r}\frac{(\nu_{p}-\nu_s)}{\nu_{s}}G
I_pI_s.\en

\section{heat transfer efficiency}

We next need to determine the part of this heat that is actually
available as acoustic energy.  This is parameterized by $\eta$
which depends on how much of the Raman energy absorbed by a spore
can be transferred to the surrounding gas in the optoacoustic cell
on the time scale of acoustic modes.  As we noted in the previous
section the heat is lost from the spore through radiation and
through collisions with gas molecules.

We first calculate the collisional energy transfer rate. Assuming
a spherical particle of radius $r_s$, the rate of collisions $N_c$
of gas molecules with the surface of the particle is determined by
\bn P_0\times 4\pi r_s^2=N_c\times 2m_g u_g\en
with mean velocity $u_g$ of the gas molecules given by $m_g\langle
u_g\rangle =\sqrt{3m_gkT_0}$; at standard temperature and pressure
(STP), $N_c\sim 4\times 10^{17}$ collisions/second. The
temperature increase of the spore due to Raman absorption is given
by
\bn c_v\Delta T=f_RN_a\hbar(\nu_p-\nu_s)\label{endeposit}\en
where $N_a$ is the Avogadro number and $c_v$ the molar specific
heat.   Most of the molecules in a bacterial spore are usually
water (for example in anthrax spores $f_R\sim 0.1$) and so we use
the specific heat of water $c_v=4.2\ {\rm J/mole/^\circ K}$. Then
for typical vibrational frequencies of $\sim 10^{14}$ Hz and Raman
active fraction in the range $f_R=0.1-0.5$, the spore will gain a
temperature of $100^{\circ}-500^{\circ}$K. Thus the
\emph{increase} in temperature $T_S$ of the spore after Raman
scattering is of the order of the initial temperature of the gas
($\sim 300^\circ$ K), and since the water and the diatomic gas
molecules have similar masses, the energy transfer in each
collision will also be of the same order $\sim kT_0\sim kT_S$.
Therefore the rate of heat transfer from the particle to the gas
is given by
\bn c_v\dot{T}_s\sim -N_c kT_s\times
\frac{N_a}{N_S}=-\frac{N_c}{N_S}RT_S, \en
and the time scale of the collisional heat loss is $\tau\sim
c_vN_S/RN_c$. The number of water molecules in a micron-size
particle is $\sim V_S\times N_a\simeq 10^{12}$. Thus $\tau\sim
10^{-5} {\rm s}\ll$ the typical acoustic modulation times of $\sim
10^{-2}$ s.

The rate of heat loss by radiation is given by
 \bn \sigma_B\times
T_s^4\times 4\pi r_s^2\ {\rm W}\en
where $\sigma_B=5.670\times 10^{-8}\ {\rm W m^{-2}K^{-4}}$ is the
Stefan-Boltzmann constant. At the upper limiting temperature of
$1000^\circ$K for a micron-size particle the radiative flux is
$\sim 10^{-7} {\rm W}$. The energy deposited in the spore is given
by $N_S\hbar(\nu_p-\nu_s)\simeq 10^{-8}$ J, so the time scale of
radiative loss is of the order of $0.1$ s, which is much slower
than the collisional loss rate.  Therefore we can assume that on
the time scale of the acoustic waves, essentially all the heat
from the spores is transferred to the gas through collisions, and
so we can take $\eta=1$.

\section{Acoustic signals}

The absorbed heat will cause a pressure wave which will radiate
out from each spore.  Considering that the time for a laser pulse
to traverse the optoacoustic device is very short compared to the
time scale of sound propagation, the heat absorption and
subsequent release is simultaneous for all the spores.  So the
wavefronts from each spore will be in sync and will add
constructively to cause a macroscopic expansion/contraction of gas
in the optoacoustic device. This allows us to describe the
acoustic waves in the tube by a \emph{macroscopic} wave equation
for which $H$ serves as the source.

When the density of the spores is sufficiently high to give a
statistically uniform distribution, the wavefronts from all of
them will merge to yield acoustic waves which bear the symmetry of
the beam profile, typically cylindrical; on the other hand if the
density is such that the number of spores in the tube is of order
unity, the waves will bear the symmetry of the spore which is
roughly spherical. For an estimate of signal and noise we need not
delve into the specific nature of the acoustic modes, except to
note that only the lowest modes are significantly excited.

There is no net transport of matter for weak acoustic modes, so
the waves can be described by relative displacement of
infinitesimal volume segments in the medium; we denote the
displacement from equilibrium of the air at distance $r$ from the
source by $\xi(\vec{r},t)$ and the associated variation from the
equilibrium pressure by $p(\vec{r},t)=P(\vec{r},t)-P_0$. The
propagation of the disturbance is then described by
\bn\label{geneqn1}
\frac{d^2\xi}{dt^2}+\Gamma\frac{d\xi}{dt}-c^2\nabla^2
\xi=\frac{1}{V\rho_0}[F(\vec{r},t)+R(t)]\en
The sound velocity in the medium is $c^2=P_0\gamma/\rho_0$ with
$\rho_0\rightarrow$ equilibrium density, $V\rightarrow$ the volume
of the gas cell and $\gamma=C_p/C_v\rightarrow$ the ratio of
specific heat at constant pressure to that at constant volume. The
macroscopic damping rate of the acoustic modes is denoted by
$\Gamma$ and the external driving force is $F(\vec{r},t)$. We have
added a random force $R(t)$ to allow a \emph{Langevin} treatment
of the fluctuations. The calculation of the noise in terms of the
displacement variable has the intuitive appeal that the driving
terms carry dimensions of force.

The process of heat transfer on the other hand is best described
in terms of a thermodynamic variable and we will write an
equivalent wave equation for the pressure variation $p$ which is
related to Eq.~(\ref{geneqn1}) by
\bn\label{xiandp} {\rm Newton's\ law}\ \
\rho_0\frac{\partial^2\xi}{\partial t^2}=-\nabla p\h{2mm} {\rm
and}\h{2mm} p=-\rho_0 c^2\nabla\xi. \en
Before we write the equation we determine the form of the driving
force.  For fixed  total volume of the gas in the cell, basic
thermodynamics relates the change in heat to the pressure
differential
\bn dQ=dE=\frac{VC_v}{R}dP\h{2mm}\Rightarrow\h{2mm}
dp=\frac{(\gamma-1)}{V}dQ\en
Earlier we defined $H$ as the heat generated per second per unit
volume of the gas due to the incident light so that $dH\equiv
\frac{1}{V}dQ/dt$, it carries dimensions of power/volume $=$
pressure/time. This heat drives the acoustic waves and we can now
write the wave equation for the pressure variations
\bn\label{gaseqn2} \frac{\partial^2 p }{\partial t^2
}+\Gamma\frac{\partial p }{\partial t }- c^2\nabla^2
p=(\gamma-1)\frac{\partial H(\vec{r},t)}{\partial t}+R_p(t). \en
Here the random term $R_p(t)$ is not a force, but it can be
derived from the force $R(t)$ in the displacement equation.  The
first step to the solution of the wave equation is to eliminate
the spatial derivative by doing an expansion in terms of the
normal modes of the corresponding homogenous equation
\bn\label{modexp} p(\vec{r},t)=\sum_j A_j(t) p_j(\vec{r});\h{5mm}
\left(\nabla^2+\frac{\omega_j^2}{c^2}\right) p_j(\vec{r})=0\en
where the mode functions $p_j(\vec{r})$ and the mode frequencies
$\omega_j$ are determined by the symmetry of the system and the
boundary condition of vanishing velocity at the cell-walls. The
displacement $\xi(\vec{r},t)$ has an identical expansion with the
\emph{same} frequencies but different mode functions
$\xi_j(\vec{r})$ and amplitudes $q_j(t)$. The modes are orthogonal
and normalized to
\bn \int dV p_i(\vec{r})p_j(\vec{r})=V\delta_{ij}.\en

\section{Signal, noise and NEP}

The signal strength is determined by the mode amplitudes of the
pressure variations that satisfy
\bn\label{eqnA} \h{-1mm}\frac{\partial^2 A_j }{\partial t^2
}+\h{-1mm}\Gamma_s \frac{\partial A_j }{\partial t
}+\h{-1mm}\omega_j^2 A_j
=\h{-1mm}\frac{(\gamma-1)}{V}\frac{\partial}{\partial
t}\h{-1mm}\int\h{-1mm} dV p_j(\vec{r})H(\vec{r},t) \en
The signal damping rate $\Gamma_s$, which determines the profile
of the acoustic signal, is determined primarily by macroscopic
thermal and viscous losses. After doing a Fourier transform,
$A_j(t)=\int d\omega e^{-i\omega t}A_j(\omega)$ etc., we obtain an
expression for the mode amplitudes in the frequency domain
\bn\label{sigspec} A_j(\omega)=\frac{i\omega(\gamma-1)\int dV
p_j(\vec{r})H(\vec{r},\omega)}{V(\omega_j^2-\omega^2+i\omega\Gamma_s)}.
\en

To calculate the noise we use the normal mode expansion of the
displacement in Eq.~(\ref{geneqn1}) and write the resultant second
order Langevin wave equation for the mode amplitudes $q_i(t)$ as
two coupled linear equations
\bn \frac{du_j}{dt}+\Gamma_n u_j=\frac{1}{V\rho_0}[F'(t)+R(t)];
\h{5mm} \frac{dq_j}{dt}=u_j.\en
with $F'(t)=-\omega_j^2q_j(t)+\frac{1}{V}\int dV
\xi_j(\vec{r})F(\vec{r},t)$. The damping rate $\Gamma_n$ of the
noise amplitudes is mainly determined by the characteristics of
the detector and should be distinguished from the signal damping.
A time-frequency Fourier transform of the first of the two
equations, along with the property of the noise that $\langle
R(t)\rangle=0$, determines the power spectrum of the velocity
distribution
\bn \langle u_j(\omega)u_j(\omega')\rangle=\frac{\langle
R(\omega)R(\omega')\rangle}{(V\rho_0)^2(\Gamma_n^2+\omega^2)}.\en
Assuming white noise, the noise is delta-correlated $\langle
R(\omega)R(\omega')\rangle\propto\delta(\omega-\omega')$. The
Weiner-Khintchine theorem gives the \emph{correlations} for the
noise force in the time domain, which define the diffusion
coefficient $D$,
\bn \langle R(t)R(t')\rangle=2D\delta(t-t')\en
An inverse Fourier transform and the \emph{equipartition theorem}
for each mode in thermal equilibrium, $\rho_0 V\langle
u_j^2(t)\rangle=kT$, yields the diffusion coefficient
\bn\label{diffcoeff} \langle u_i(t)u_i(t')\rangle
=\frac{D}{(\rho_0 V)^2\Gamma_n}e^{-\Gamma_n(t-t')}
\h{3mm}\Rightarrow D=\rho_0 V\Gamma_n kT\en
Now we take a Fourier transform of the second order equation in
the displacement mode amplitudes
\bn
q_j(\omega)=\frac{F'(\omega)+R(\omega)}{m(\omega_0^2-\omega^2+i\omega\Gamma_n)}\en
and use expression (\ref{diffcoeff}) for the diffusion coefficient
and the relation between $\xi$ and $p$ in Eq.~(\ref{xiandp})
 \bn p=-\rho_0 c^2\nabla \xi \Rightarrow
A_j=-i\rho_0 c\omega_jq_j\en
to get the noise spectrum of pressure-wave amplitude
\bn\label{noisespec} \langle
|A_{jn}|^2(\omega)\rangle=\frac{\rho_0 c^2\omega_j^2\Gamma_n
kT}{V[(\omega_j^2-\omega^2)^2+(\omega\Gamma_n)^2]}. \en

The noise equivalent power (NEP) is defined to be the input power
necessary to produce a signal  amplitude equal to the noise
amplitude \cite{Kreuzer}. In order to estimate the NEP we note
that the signal is typically generated in the lowest mode
$\omega_0=0$ corresponding to an uniform excitation of the gas.
The noise is also dominated by the lowest modes for weak
excitations.  For the very lowest mode the noise amplitude
vanishes, so we take the amplitude of the next higher one
$\omega_1$ as a measure of the magnitude of the noise. Thus the
noise equivalent power is approximately given by the value
(indicated by subscript NEP) of the input power $VH$ for which
$|A_0(\omega)|^2\simeq |A_{1n}(\omega)|^2$. Using the expressions
from Eqs.~(\ref{sigspec}) and (\ref{noisespec})  we get
\bn
 |VH_{NEP}(\omega)|^2 &\simeq &\frac{V\Gamma_n\rho_0
c^2kT(\omega^2+\Gamma_s^2)}{\omega_1^2(\gamma-1)^2}\n\en
where we also used the fact that acoustic modulation frequency is
typically much smaller than the mode frequencies
$\omega\ll\omega_{j\neq 0}$.

\section{Minimum detectable density}

The minimum detectable number density of spores $\rho_s$ can be
estimated by setting the input power density $H=H_{NEQ}$, the
noise equivalent power density in Eq.~(\ref{power}). For the
purpose of numerical estimation we will use the specific example
of an anthrax spore for which the Raman active molecule is
dipcolinic acid (DPA) which constitutes $17\%$ of the weight, the
rest being mainly water. Since some of the properties of DPA are
not easily available, for those we use the values for benzene an
organic molecule similar in structure.

The  dimensions of gas cell are optimized for the maximum
interaction with the input beams, for a typical cell length of
$l=0.1$ m, the optimum radius for focussing the beams into the
cell is given by $r=\sqrt{\lambda l/\pi}\simeq 1.4\times 10^{-4}$
if we use a wavelength in the visible range $\lambda\simeq 7\times
10^{8}$  so that the cell volume $V\simeq \pi r^2l\simeq 10^{-8}\
m^3$. We will assume that by placing multiple microphones all of
the generated signal can be detected, but this can be easily
relaxed by including a geometrical factor which involves the ratio
of detector surface area to the cell surface area.

At standard temperature and pressure, $T_0=300^\circ\ {\rm K}$ and
$P_0=1\ {\rm atm.}$ and $\rho_0=1.3 {\rm kg\ m^{-3}}$; for air
$\gamma=1.4$ being mainly diatomic gases. We take the dominant
noise mode to be at the resonant frequency of the detector,
typically a condenser microphone, $\omega_{1}\simeq \omega_m$, and
a reasonable value\cite{Kreuzer} is $4\times 10^{4} {\rm Hz}$, the
damping of the mode has the same order of magnitude
$\Gamma_n=5\times 10^{4} {\rm Hz}$ assuming a quality factor
${\cal Q}_m=\omega/\Gamma=0.8$. The optimum modulation frequency
is determined by the thermal damping time $\omega\simeq
\Gamma_s\simeq 100 s^{-1}$.  Using these values we find
\bn
 H_{NEP}(\omega) &\simeq & 5\times 10^{-5}
{\rm W m^{-3}s^{1/2}}\n\en

In order to calculate the Raman gain factor $G$,  we use optical
frequencies for the input beams $\nu_{p(s)}\sim 2\pi\times 4\times
10^{14}{\rm s^{-1}}$.  The number density of DPA molecules in an
anthrax spore is $4\times 10^{26}$ molecules  m$^{-3}$.  We use
the Raman cross-section for Benzene
$\frac{d\sigma}{d\Omega}=32.5\times 10^{-34} {\rm
m^2/sr/molecule}$. A typical linewidth of a Raman line for Benzene
is about $\delta \nu \sim 6.45 \times 10^{10}{\rm Hz}$. Using
these values we get for the gain factor
\bn G=8.5 \times 10^{-12}{\rm mW^{-1}}\en

Finally noting that an anthrax spore has dimensions $1\times
2\times 1\ {\rm \mu m}^3$ we can use Eq.~(\ref{power}) to find the
minimum detectable number density of anthrax spores in air
\bn \rho_S\simeq \frac{H_{NEP}\sqrt{\Gamma_s}}{\eta\times
H_R\times V_S} =\frac{3\times 10^{25}}{I_pI_s}{\rm m^{-3}}\n\en
where we multiplied by the square root of the width of the
acoustic signal since the NEP is not really ``power" but
power/$\sqrt{\rm frequency}$.
With powerful lasers available today intensities of
$I_{p(s)}\simeq 10^{12}{\rm Wm^{-2}}$ are reasonable, in which
case one would get $\rho_s\simeq 30 {\rm m^{-3}}$.

Note that so one cannot indefinitely increase the laser
intensities because at intensities of $10^{16}{\rm Wm^{-2}}$ there
is cascade breakdown \cite{Kroll} of air at STP. While cooling the
gas cell seems to be easy way to reduce thermal noise, due to
condensation effects temperatures cannot be lowered too much. Also
it is much simpler and realistic to do conduct tests for
atmospheric contaminants at room temperatures.

\section{Conclusion}

In conclusion, we developed a simple thermodynamic model to
describe the heat transfer mechanisms and generation of acoustic
waves in photoacoustic Raman spectroscopy for detection of trace
amounts of small solid particles suspended in air. As it has been
pointed out since its first application \cite{Barrett} PARS has
several advantages over other stimulated Raman techniques: it is
not plagued by a non-resonant background and there is no necessity
for phase-matching, two very crucial issues in Coherent
Anti-Stokes Raman Spectroscopy (CARS) for example, also PARS
directly measures the deposited Raman energy instead of a
fractional energy change. The noise in PARS arises primarily from
thermal fluctuations which may be reduced by controlling
macroscopic parameters, unlike in other Raman process where the
noise arises from microscopic quantum fluctuations harder to
control. Even though thermal noise in a gas is well documented, we
applied Langevin methods to calculate it for our particular
physical situation; a reason for doing that is that it can be
immediately applied to include other sources of noises in more
detailed studies in the future.

We applied our model to the specific case of anthrax spores and
showed that as few as $100$ spores in a volume of one cubic meter
could be detected if thermal noise is the main limitation in the
system.  This promises potential development of a PARS-based
method to detect such harmful contaminants in real time with
relatively simple means  at room temperatures, something that is
certainly of very strong practical and immediate interest.

\section*{Acknowledgments}

We thank S. E. Harris for useful discussions and gratefully
acknowledge the support from the Office of Naval Research, the
Air Force Research Laboratory (Rome, NY), Defense Advanced
Research Projects Agency-QuIST, Texas A$\&$M University
Telecommunication and Information Task Force (TITF) Initiative,
and the Robert A. Welch Foundation.

\end{document}